\documentclass[pra,twocolumn,showpacs,superscriptaddress,amsfonts,amsmath,floatfix]{revtex4}
\usepackage{graphicx}
\newcommand{\Journal}[4]{#1 {\bf #2}, #3 (#4)}%
\newcommand{\PR}{Phys. Rev.}
\newcommand{\PRL}{Phys. Rev. Lett.}
\newcommand{\PRA}{Phys. Rev. A}

\newcommand{\PRE}{Phys. Rev. E}
\newcommand{\JMP}{J. Math. Phys.}

\newcommand{\Science}{Science}

\newcommand{\A}{\text{A}}
\newcommand{\B}{\text{B}}
\newcommand{\D}{\text{D}}
\newcommand{\F}{\text{F}}
\renewcommand{\AA}{\text{AA}}
\newcommand{\AB}{\text{AB}}
\newcommand{\BB}{\text{BB}}
\newcommand{\nA}{n_\text{A}}
\newcommand{\nB}{n_\text{B}}
\newcommand{\NA}{{N_\text{A}}}
\newcommand{\NB}{{N_\text{B}}}
\newcommand{\CS}{\text{CS}}
\newcommand{\FTG}{\text{FTG}}
\newcommand{\pure}{\text{pure}}
\newcommand{\sgn}{\mathop{\rm sgn}\nolimits}

\begin{document}
\title{Ground state of a mixture of two bosonic Calogero-Sutherland gases with strong odd-wave interspecies attraction}
\author{M. D. Girardeau}
\email{girardeau@optics.arizona.edu}
\affiliation{College of Optical Sciences, University of Arizona, Tucson, AZ 85721, USA}
\author{G.E. Astrakharchik}
\email{astrakharchik@mail.ru}
\affiliation{Departament de F\'{\i}sica i Enginyeria Nuclear, Campus Nord B4, Universitat Polit\`ecnica de Catalunya, E-08034 Barcelona, Spain}
\date{\today}
\begin{abstract}
A model of two Calogero-Sutherland Bose gases A and B with strong odd-wave AB attractions induced by a $p$-wave AB Feshbach resonance is studied. The ground state wave function is found analytically by a Bose-Bose duality mapping, which permits to accurately determine static physical properties by a Monte Carlo method. The condensation of particles or particle pairs (molecules) is tested by analyzing the presence of the off-diagonal long-range order in one- or two- body density matrices. The $p$-wave symmetry of AB interaction makes possible quasi-condensation of type A particles at the Fermi momentum of the B component. The zero-temperature phase diagram is drawn in terms of densities and interaction strengths.
\end{abstract}
\maketitle
\section{Introduction}
Strong interatomic interactions and correlations occur in ultracold gases confined in de Broglie waveguides with transverse trapping so tight that the atomic dynamics is essentially one-dimensional (1D) \cite{Ols98}, with confinement-induced resonances \cite{Ols98,GraBlu04} allowing Feshbach resonance tuning \cite{Rob01} of the effective 1D interactions to very large values. This has led to experimental verification \cite{Par04Kin04,Kin05,Kin06} of the fermionization of bosonic ultracold vapors in such geometries predicted by the Fermi-Bose (FB) mapping method \cite{Gir60Gir65}, an exact mapping of a 1D gas of bosons with  point hard core repulsions, the ``Tonks-Girardeau'' (TG) gas, to an \emph{ideal} spin-aligned Fermi gas. The ``fermionic Tonks-Girardeau'' (FTG) gas \cite{GirOls04,GirNguOls04}, a 1D spin-aligned Fermi gas with very strong \emph{attractive} interactions, can be realized by a 3D $p$-wave Feshbach resonance as, e.g., in ultracold $^{40}$K vapor \cite{Tik04}. It was shown recently \cite{Gir09_3} that a mixture of two 1D ideal \emph{Bose} gases A and B with a strongly attractive AB interaction of the same FTG form is exactly solvable by a Bose-Bose duality mapping to a mixture of two ideal Bose gases with no AB interaction. Reduced density matrices of all orders were shown to be reducible to 1D integrals, and it was found that the off-diagonal elements of the one- and two-particle density matrices have unusual behavior: The strong AB attraction destroys the ground state Bose-Einstein condensation (BEC) and single-particle off-diagonal long-range order (ODLRO) of both components A and B, and it induces both AA and BB pairing manifested in superconductive ODLRO of the two-particle density matrices of components A and B, although there are no AA or BB interactions. Furthermore, there is no AB pair ODLRO in spite of the strong AB attractions. It was also shown that if the AB attraction is a finite odd-wave attraction rather than the infinite FTG limit, and there is also a repulsive even-wave AB interaction of Lieb-Liniger (LL) $\delta$-function form \cite{LieLin63}, then there is a quantum phase transition as the coupling constants are varied, between a phase in which there are no AB contact nodes and only the repulsive LL interaction acts, and another phase in which there are AB contact nodes and only the attractive FTG-like interaction acts.

The Calogero-Sutherland (CS) model \cite{C69S71} of a 1D Bose or Fermi gas with interparticle interaction potential $\lambda(\lambda-1)/x_{ij}^2$ is analytically solvable for all values of the coupling constant $\lambda$. Here we will show that a mixture of two CS Bose gases A and B with an AB interaction of FTG form is also analytically solvable by the same Bose-Bose duality mapping described in the previous paragraph, which in this case maps the system to a mixture of two CS gases with no AB interaction. This has the great advantage that
the single-component CS model reduces to previously known results in several limits, thus providing an exact interpolation between those limits: ideal Bose gas for $\lambda = 0$, TG gas for $\lambda = 1$, and the limit of a classical crystal $\lambda\to\infty$. For the mixture, the ground state can be found in closed form even if components A and B have different CS coupling constants $\lambda_\A$ and $\lambda_\B$.
The ground state will be found in explicit analytical form and the one- and two-particle density matrices will be determined numerically by a Monte Carlo method.
\section{Single-component Calogero-Sutherland model\label{sec:Single-component Calogero-Sutherland model}}
Before studying the properties of mixtures it is useful to recall some properties of a single-component CS system. For $N$ particles of mass $m$ in a periodic box of size $L$, the CS interaction potential is
\begin{eqnarray}
V^{\CS}(x) = \frac{\pi^2\hbar^2}{mL^2}\frac{\lambda(\lambda-1)}{\sin^2 (\pi x /L)}\ ,
\label{eq:VCS}
\end{eqnarray}
where $\lambda\ge 0$ is the interaction parameter. This interaction potential satisfies periodic boundary conditions and can be obtained by evaluating the $L$-periodic extension of an inverse square potential:
$V^{\CS}(x) = \sum_{j=-\infty}^\infty \hbar^2\lambda(\lambda-1) / [m(x+jL)^2]$.
Another physical interpretation of the sine function in Eq.~(\ref{eq:VCS}) is that particles stay on a ring of diameter $L$ but the two-particle interaction corresponds to the chord distance $L\sin(\pi x/L)$. In the thermodynamic limit the interaction potential reduces to an inverse square potential. Such a potential is very special as it scales as the kinetic energy and there is no any other length scale in the system different from density. Properties of the system are then governed by the interaction parameter  $\lambda$. The bosonic ground-state wave function of particles interacting with the potential Eq.~(\ref{eq:VCS}) can be written explicitly:
\begin{eqnarray}\label{psiCS}
\psi^{\CS}(x_1,...,x_N) = \prod\limits_{i<j}^N \left|\sin \frac{\pi(x_i-x_j)}{L}\right|^\lambda\ .
\label{eq:wfCS}
\end{eqnarray}
An important observation is that the tails of the interaction potentials vanish for two special values $\lambda = 0$ and $\lambda = 1$. In the first case the ground state wave function reduces to a constant which is the solution for an ideal Bose gas, while in the second case the solution~(\ref{eq:wfCS}) coincides with the absolute value of the wave function of an ideal Fermi gas and corresponds to the TG gas \cite{Note1}.

Although at present there are no realizations of the CS system, this model describes many important physical regimes that can be reached in other one-dimensional systems. Furthermore, the long-range properties are expected to be similar. In order to see that we note that the long-range wave function of a one-dimensional system can be deduced from zero-point motion of the phonons. The ground state has the form \cite{Reatto} $\psi(x_1,...,x_N)=\psi_{\text{short}}(x_1,...,x_N)\psi_{\text{phon}}(x_1,...,x_N)$ where $\psi_{\text{short}}$ describes the short-range behavior and
\begin{eqnarray}
\psi_{\text{phon}}(x_1,...,x_N) = \prod\limits_{i<j}^N \left|\sin\frac{\pi(x_i-x_j)}{L}\right|^{\alpha/2}
\label{eq:wfphonons}
\end{eqnarray}
for large separations between particles. Here $\alpha$ is a constant related to the speed of sound. Comparing (\ref{eq:wfphonons}) to (\ref{eq:wfCS}) we see that actually the CS wave function retains the structure typical of phonons even at short distances.

The presence of phonons in a one-dimensional system permits construction of an effective Luttinger liquid model, which is based on the assumption that the excitation spectrum is linear in momentum $E_k = \hbar|k|/2mc$, where $c$ is the speed of sound. The Luttinger liquid yields the long-range behavior of the correlation functions and leads to some important conclusions on the generic behavior of one-dimensional systems, such as the absence of Bose-Einstein condensation and absence of crystalline order. Within this approach properties are defined by one governing parameter, the Luttinger constant
$K=v_F/2c$ with $v_F = \pi\hbar n/m$ being the Fermi velocity. The speed of sound in the CS model can be calculated from the compressibility $mc^2 = n\partial\mu/\partial n$ (as usual $\mu=\partial E/\partial N$ is the chemical potential) and leads to a very simple relation $K = 1/\lambda$. This means, in particular, that the CS model has the same Luttinger parameter as the $\delta$-interacting LL gas in the region $0<\lambda<1$. Values of $\lambda$ slightly larger than unity correspond to the ``super-Tonks-Girardeau'' regime. Finally, $\lambda>2$ correspond to a quasi-crystalline region, with diverging peaks in the static structure factor. From this point of view the CS model describes long-range properties in very different physical regimes, ranging from the ideal Bose gas to a classical crystal. For a detailed study of these various regimes see \cite{AGLS06}. In the regime $\lambda\le 1/2$, in addition to the well-behaved solution (\ref{psiCS}) the CS model possesses singular solutions associated with ``fall to the center'' \cite{Landau,Note1.5}. Nevertheless, the state (\ref{psiCS}) remains well-behaved; we conjecture that for $0<\lambda<1/2$ it is the lowest gas-like state. The long-range properties of this state agree with those of the LL gas in the whole regime $0<\lambda<1$.

\section{FTG interaction}
The FTG gas is a spin-aligned 1D Fermi gas with infinitely strongly attractive zero-range odd-wave interaction induced by a $p$-wave Feshbach resonance. It is the infinite 1D scattering length limit $a_{1\D}\to -\infty$ of a 1D Fermi gas with zero-range attractive interactions leading to a 1D scattering length defined in terms of the ratio of the derivative $\Psi^{'}$ of the wave function to its value at contact:
$\Psi(x_{jk}=0+)=-\Psi(x_{jk}=0-)= -a_{1\D}\Psi^{'}(x_{jk}=0\pm)$ where the prime denotes the derivative with respect to $x_{jk}$
\cite{GraBlu04,GirOls04,GirNguOls04}.
The FTG limit $a_{1\D}\to -\infty$ corresponds to a 1D zero-energy odd-wave scattering resonance reachable by Feshbach resonance tuning to a 1D odd-wave confinement-induced resonance \cite{Rob01,Ols98,GraBlu04}. An AB interaction of this form was used in Ref.~\cite{Gir09_3} for a mixture of two 1D ideal Bose gases A and B, and here we will use it for the AB interaction between two CS Bose gases A and B; Ref.~\cite{Gir09_3} can be consulted for more details of the interaction and its effect.
\section{Two-component mixture and mapping solution}
Consider now a mixture of two CS gases A and B with an AB interaction of FTG form and no longitudinal trap potential, with periodic boundary conditions of periodicity length $L$. The Hamiltonian is
\begin{eqnarray}\label{H}
&&\hat{H}=\sum_{i=1}^\NA\frac{-\hbar^2}{2m_A}\frac{\partial^2}{\partial x_i^2}+\sum_{1\le i<j\le \NA}V_{\A}(x_i-x_j)\nonumber\\
&&+\sum_{i=1}^\NB\frac{-\hbar^2}{2m_B}\frac{\partial^2}{\partial y_i^2}+\sum_{1\le i<j\le \NB}V_{\B}(y_i-y_j)\nonumber\\
&&+\sum_{i=1}^\NA\sum_{j=1}^\NB\hat{v}_{\FTG}(x_i-y_j)\ ,
\end{eqnarray}
where $(x_1,\cdots,x_\NA)$ are the particle coordinates for component A, $(y_1,\cdots,y_\NB)$ are those for component B, $V_{\A}(x)$ and $V_{\B}(x)$ are the previously-defined CS interaction $V^{\CS}(x)$ with CS coupling constants $\lambda_\A$ and $\lambda_\B$, and $\hat{v}_{\FTG}(x)$ is the FTG interaction described in \cite{GraBlu04,GirOls04,GirNguOls04,Gir09_3} and Sec. III above. The exact ground state of the model of \cite{Gir09_3} was found with the aid of a Bose-Bose duality mapping function \cite{Note2}
\begin{equation}\label{M}
M(x_1,\cdots,x_\NA;y_1,\cdots,y_\NB)
=\prod_{i=1}^\NA\prod_{j=1}^\NB\sgn(x_i-y_j)\ ,
\end{equation}
where the sign function $\sgn(x)$ is $+1\ (-1)$ if $x>0\ (x<0)$, and the same mapping can be used to obtain the exact solution of the present model. Let $\Psi_0$ be the exact ground state of our mixture of two CS gases A and B with FTG AB interactions, which we wish to determine. As in \cite{Gir09_3}, mapping by $M$ removes the interaction between components A and B, in the sense that $\Psi_0=\Psi_{\text{M}0}M$ where $\Psi_{\text{M}0}$, the ground state of a mixture of two CS gases A and B with no AB interactions, is the product $\Psi_{\text{M}0}=\Psi_{0\A}\Psi_{0\B}$ of the ground states of CS gases A and B. The explicit expression for the ground state wave function is
\begin{eqnarray}
\nonumber
\Psi_0 &=& \prod_{i=1}^\NA\prod_{j>i}^\NA\prod_{k=1}^\NB\prod_{l>k}^\NB |\sin\pi(x_i-x_j)/L|^{\lambda_\A}\\
&&
~~~~~~~~|\sin\pi(y_k-y_l)/L|^{\lambda_\B}
\sgn(x_i-y_k)\ .
\label{eq:psi_0}
\end{eqnarray}
Since $[M(x_1,\cdots,x_\NA;y_1,\cdots,y_\NB)]^2=1$, all diagonal density matrix elements of the ground state $\Psi_0$ with FTG AB interactions are the same as those of the model ground state $\Psi_{\text{M}0}$ which has no AB interactions. In particular, the A and B component densities in the present case (no trap potential) are trivial constants, $n_{\A}(x)=\NA/L$ and $n_{\B}(y)=\NB/L$ where $L$ is the period of the periodic boundary conditions. Furthermore, the AA and BB pair distribution functions are just those of single-component CS gases with coupling constants $\lambda_{\A}$ and $\lambda_{\B}$, and the AB pair distribution function is just a constant $n_{\A}n_{\B}$.
\section{One-particle density matrices and momentum distributions}
Next consider the off-diagonal elements of the single-particle density matrices $\rho_{1A}(x,x')$ and $\rho_{1B}(y,y')$ of components A and B. One has
\begin{widetext}
\begin{eqnarray}\label{rho1A}
&&\rho_{1\A}(x,x')=\NA L^{-\NA-\NB}\int\Psi_0(x,x_2,\cdots,x_{\NA};Y)\Psi_0(x',x_2,
\cdots,x_{N_\A};Y)dx_2\cdots dx_{\NA}dY\nonumber\\
&&=\int\Psi_{0\A}(x,x_2,\cdots,x_{\NA})\Psi_{0\A}(x',x_2,\cdots,x_{\NA})
\Psi_{0\B}^2(y_1,\cdots,y_{\NB})\prod_{j=1}^{\NB}\sgn(x-y_j)\sgn(x'-y_j)
dx_2\cdots dx_{\NA}dY\nonumber\\
&&=\rho_{1\A}^{\pure}(x,x')
\int\Psi_{0\B}^2(y_1,\cdots,y_{\NB})\prod_{j=1}^{\NB}\sgn(x-y_j)\sgn(x'-y_j)dY\ ,
\end{eqnarray}
where $Y=(y_1,\cdots,y_{N_\B})$ and $\rho_{1\A}^{\pure}(x,x')$ is the single-particle density matrix of the pure (single-component) CS gas. The formula for $\rho_{1\B}(y,y')$ is the same, with obvious interchanges of A with B and $x$ with $y$.
%
\section{Two-particle density matrices}
By generalization of (\ref{rho1A}) and the mapping used in \cite{GirMin06,Gir09_3} the two-particle A-component density matrix $\rho_{2\AA}(x_1,x_2;x_1',x_2')$ is
\begin{eqnarray}\label{rho2AA}
&&\rho_{2\AA}(x_{1},x_{2};x_{1}',x_{2}')=\NA(\NA-1)L^{-(N_\A+N_\B)}
\int\Psi_0(x_1,x_2,x_3,\cdots,x_{\NA};Y)
\Psi_0(x_1',x_2',x_3,\cdots,x_{\NA};Y)dx_3\cdots dx_{\NA}dY\nonumber\\
&=&\int\Psi_{0\A}(x_1,\cdots,x_{\NA})\Psi_{0\A}(x_1',x_2',x_3,\cdots,x_{\NA})
\Psi_{0\B}^2(y_1,\cdots,y_{\NB})\nonumber\\
&\times&\prod_{j=1}^{\NB}
\sgn(x_1-y_j)\sgn(x_1'-y_j)\sgn(x_2-y_j)\sgn(x_2'-y_j)dx_3\cdots dx_{\NA}dY\nonumber\\
&=&\rho_{2\AA}^{\pure}(x_{1},x_{2};x_{1}',x_{2}')\int\Psi_{0\B}^2(y_1,\cdots,y_{\NB})
\prod_{j=1}^{\NB}\sgn(x_1-y_j)\sgn(x_1'-y_j)\sgn(x_2-y_j)\sgn(x_2'-y_j)dY\ ,
\end{eqnarray}
where $\rho_{2\AA}^{\pure}(x_{1},x_{2};x_{1}',x_{2}')$ is the two-particle density matrix of the pure CS gas,
and the formula for $\rho_{2\BB}$ differs only by the obvious interchanges. The AB-pair density matrix is
\begin{eqnarray}\label{rho2AB}
&&\rho_{2\AB}(x,y;x',y')=\NA\NB L^{-(\NA+\NB)}
\int\Psi_0(x,x_2,x_3,\cdots,x_{\NA};y,y_2,\cdots,y_{\NB})
\Psi_0(x',x_2,\cdots,x_{\NA};y',y_2,\cdots,y_{\NB})\nonumber\\
&\times& dx_2\cdots dx_{\NA}dy_2\cdots dy_{\NB}=
\int\Psi_{0\A}(x,x_2,\cdots,x_{\NA})\Psi_{0\B}(y,y_2,\cdots,y_{\NB})
\Psi_{0\A}(x',x_2,\cdots,x_{\NA})\Psi_{0\B}(y',y_2,\cdots,y_{\NB})\nonumber\\
&\times&\sgn(x-y)\sgn(x'-y')
\left[\prod_{j=2}^{\NB}\sgn(x-y_j)\sgn(x'-y_j)\right]\nonumber\\
&\times&
\left[\prod_{i=2}^{\NA}\sgn(x_i-y)\sgn(x_i-y')\right]
dx_2\cdots dx_{\NA}dy_2\cdots dy_{\NB}\ .
\end{eqnarray}
\end{widetext}

\section{Monte Carlo technique}
In order to obtain numerically the correlation functions we resort to Monte Carlo methods. Such methods are very efficient for evaluation of multidimensional integrals which in our case correspond to operator averages
$\langle \hat A \rangle$ = $\int...\int A(x_1, ..., x_{N_\A},y_1,...,y_{N_\B})$ $|\Psi_0(x_1, ..., x_{N_\A},y_1,...,y_{N_\B})|^2$ $dx_1$ ... $dx_{N_\A}$ $dy_1$ ... $dy_{N_\B}$, where $\hat A$ is some operator of interest. We obtained an explicit expression for the ground state wave function $\Psi_0$, as given by Eq.~(\ref{eq:psi_0}), which significantly simplifies the numerical calculations. The basic idea of the method is to generate a sequence of points ${\bf R}_i$ in the phase space ${\bf R} = \{x_1, ..., x_{N_\A},y_1,...,y_{N_\B}\}$ according to the probability distribution $|\Psi_0({\bf R})|^2$. Then the operator average is approximated by an average over ${\bf R}_i$ as $\langle A\rangle \approx \sum_{i=N_{measure}} A({\bf R}_i)/N_{measure}$. The statistical error of the estimation is kept under control and can be reduced by increasing the series of measurements. We follow Metropolis prescription \cite{Metropolis53} for generation of the sequence of ${\bf R}_i$ such that each new point in the phase depends only on the previous point ${\bf R}_{i-1}$ (i.e. such a sequence forms a Markov chain):
\begin{itemize}
\item generate a new configuration ${\bf R}_i$
\item the move from ${\bf R}_{i-1}$ to ${\bf R}_i$ is always accepted if the wave function in the trial configuration is larger, $\Psi_0({\bf R}_i)>\Psi_0({\bf R}_{i-1})$
\item if the wave function in the trial configuration is smaller, accept the move with the probability $[\Psi_0({\bf R}_i)/\Psi_0({\bf R}_{i-1})]^2$
\end{itemize}
We generate new moves by displacing a random particle by a distance taken from a Gaussian distribution. The width of the Gaussian is adjusted in such a way that the acceptance rate is close to 50\%, apart from the singular case of vanishing $\lambda_\A$ and $\lambda_\B$, when the probability distribution degenerates $|\Psi({\bf R})|^2 = 1$ and any move is accepted. In this case we adjust the width of the Gaussian in such a way that the typical displacement is small compared to the system size.

The one-particle density matrix (\ref{rho1A}) can be easily recast in a form suitable for Monte Carlo calculation. Indeed, taking into account that in a homogeneous system $\rho_{1\A}(x,x')$ is a function of the difference of arguments and labeling this difference as $x$ one has

\begin{widetext}
\begin{eqnarray}\label{rho1A(MC)}
\frac{\rho_{1\A}(x)}{\nA} =
\frac{\int\; [\Psi_0(x_1+x,x_2,\cdots,x_{N_\A};Y)/\Psi_0(X;Y)]\; \Psi^2_0(X;Y) \; dX dY}
{\int\Psi^2_0(X;Y)\; dX dY}\ .
\end{eqnarray}
\end{widetext}
The interpretation of formula~(\ref{rho1A(MC)}) is that the ratio of values of the wave function $[\Psi_0(x_1+x,x_2,\cdots,x_{N_\A};Y)/\Psi_0(X;Y)]$ with and without a particle displaced by $x$ is averaged over the random walk distributed according to $\Psi_0^2(X,Y)$.

Most generally, the two-body particle density matrix~(\ref{rho2AA}) depends on four variables, namely, $x_1, x_2, x_1', x_2'$. The physical importance of $\rho_{2\AA}(x_{1},x_{2};x_{1}',x_{2}')$ is that it can be used for testing a possible formation of a Bose condensate of particle pairs (molecules). One should study its behavior when a pair $x_1,x_2$ is displaced by some distance $R$, i.e. $x_1' = x_1+R$ and $x_2' = x_2+R$. Then the relevant parameters are the displacement length $R$ and the size of the pair $r = x_1 - x_2 = x_1'-x_2'$.
\begin{widetext}
\begin{eqnarray}\label{rho2AA(MC)}
\frac{\rho_{2\AA}(r,R)}{\nA^2}
= \left(1-\frac{1}{\NA}\right) \frac{\int [\delta(x_2-x_1-r) \Psi_0(x_1+R,x_2+R,x_3,\cdots,x_{\NA};Y) / \Psi_0(X;Y)] \Psi_0^2(X;Y) dXdY}
{\int\Psi_0^2(X;Y) dXdY}\ .
\end{eqnarray}
\end{widetext}

The ODLRO of pairs is manifested by non-zero large-$R$ asymptotic value of the projected TBDM, defined from~(\ref{rho2AA(MC)}) by integrating out the pair size $r$\cite{Senatore,Ortiz,ABCG05}
\begin{equation}
\rho_2^P(R) = \frac{2}{N}\int dx_1 dx_2 \rho_2(x_1+R,x_2+R,x_1,x_2)\ .
\label{eq:PTBDM}
\end{equation}
If ODLRO of pairs (molecules) is present, the two-body density matrix at large separation distances $R$ reduces to the product of molecule orbitals. In Monte Carlo calculations this is tested by setting $R$ to $L/4$ in Eq.~(\ref{rho2AA(MC)}).

\section{Results}
\subsection{Mixture of ideal Bose gases with FTG interaction}
We will consider first the case of vanishing $\lambda_{\A}=\lambda_{\B}=0$. This limit describes a system consisting of a mixture of two ideal Bose gases with FTG interaction between different components. Such a system is interesting as for a single component CS system $\lambda = 0$ is the only value of the interaction parameter for which true Bose condensation exists, see Section~\ref{sec:Single-component Calogero-Sutherland model}. Instead for any finite $\lambda$ the OBDM has zero asymptotic value. As well it is convenient to start the discussion from the case of mixtures of ideal Bose gases, as this the limit where explicit analytical expressions for the one- and two-particle correlation functions are known from Ref.~\cite{Gir09_3}.

\begin{figure}
\begin{center}
\includegraphics[angle=-90,width=0.9\columnwidth]{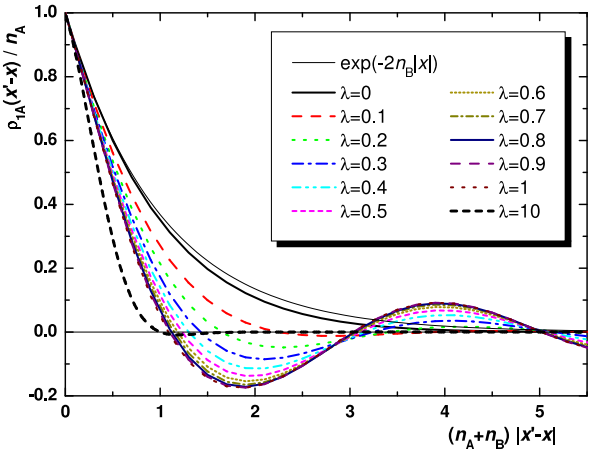}
\caption{One-body density matrix $\rho_{1\A}(x)$ in a balanced system $\NA = \NB = 10$ for different interaction strength $\lambda_\A = \lambda_\B$. Thin uppermost curve, $\lambda_\A = \lambda_\B = 0$ (mixture of two ideal Bose gases), thermodynamic limit as given by Eq.~\ref{eq:OBDM,l=0}. Thick lines, Monte Carlo results for $\NA = \NB = 10$ particles and increasing interaction strength (from upper to lower curve) $\lambda_\A = \lambda_\B = 0; 0.1; 0.2; 0.3; 0.4; 0.5; 0.6; 0.7; 0.8; 0.9; 1; 10$.}
\label{Fig:rho1(l)}
\end{center}
\end{figure}

In the absence of the other component, $\nA=0$, the OBDM is constant $\rho_{1\A}(x) = \nA$ and the Bose condensation is complete. For a finite concentration of the second component, the one-body density matrix can be obtained explicitly even in a system of a finite size and is given by $\rho_{1\A}(x) = \nA (1-2\nB|x|/\NB)^\NB$. In the thermodynamic limit the decay in $\rho_{1\A}(x)$ is exponentially fast
\begin{equation}
\rho_{1\A}(x) = \nA e^{-2 \nB|x|}
\label{eq:OBDM,l=0}
\end{equation}
and the strength of the decay depends on the density of the other component $\nB$. In particular, for zero concentration the strength of the decay is zero and the result for the ideal Bose gas is recovered. Instead, any finite concentration of the other component removes the Bose condensate in a strong sense, that is exponentially fast decay compared to power-law decay in interacting one-dimensional Bose gas (see Fig.~\ref{Fig:rho1(l)}).

The momentum distribution is related to the one-body density matrix (\ref{eq:OBDM,l=0}) by a Fourier transformation and has a Lorentzian shape \cite{Gir09_3}
\begin{equation}
\nA(k) = \frac{4\nA\nB}{4\nB^2+k^2}\ .
\label{eq:nk,l=0}
\end{equation}
An important feature of $n(k)$ in this case is that the momentum distribution is finite for zero momentum with its value defined by the ratio of the densities $\nA(0) = \nA/\nB$. Another important observation is that the high-momentum decay follows the law  $1/k^2$ contrary to $1/k^4$ decay of a Lieb-Liniger of Tonks-Girardeau gas. We will comment more on the high-momentum behavior later.

It was shown in Ref.~\cite{Gir09_3} that while FTG interaction destroys Bose condensation in each component, at the same time it induces ODLRO in TBDMs $\rho_{2\AA}$ and $\rho_{2\BB}$. Indeed, it was found that the spectral representation of the TBDM
\begin{equation}\label{eq:spectral_decomposition}
\rho_{2}(x_1,x_2;x_1',x_1') = \sum\limits_{i} \lambda_i \phi_i^*(x_1-x_2) \phi_i(x_1'-x_2')
\end{equation}
reduces in the physically interesting regime $x_1,x_2\ll x_1',x_2'$ (for example, this is the case when a pair is displaced by a large distance) to
\begin{equation}\label{eq:rho2exact}
\rho_{2\AA}(x_1,x_2;x_1',x_2') = \nA^2 e^{-2\nB|x_1-x_2|}e^{-2\nB|x_1'-x_2'|}\ .
\end{equation}
In this way the largest eigenvalue $\lambda=\NA\nA/2\nB$ is macroscopic and the corresponding eigenfunction
\begin{equation}\label{eq:phi_mol}
\phi^{\AA}(x) = \sqrt{2 \nB/L} e^{-2\nB|x|}
\end{equation}
can be interpreted as a dimer orbital. There is a BEC-BCS-like crossover from AA-pair BEC when $\nB\gg\nA$ and the range of  $\phi_{mol}^{\AA}$ is $\ll 1/\nA$, implying tightly bound AA pairs, to extended and strongly overlapping AA Cooper pairs when $\nB\ll\nA$ and the range of $\phi^{\AA}$ is $\gg 1/\nA$. The TBDM $\rho_{2\BB}$(x) of the other component exhibits similar behavior with A and B interchanged. It is interesting to note that largest eigenvalue $\lambda=\NA\nA/2\nB$ in a balanced $N_\A = N_\B = N$ system reduces to $N/2$, i.e. all $N/2$ dimers are condensed. In an unbalanced system the range of a wave function of $\AA$ dimer (\ref{eq:phi_mol}) is governed by the density $\nB$ of the other component. The eigenfunctions have to be normalized to unity $\int_{-L}^{L} |\phi(x)|^2\;Ldx = 1$ and as a result the normalization constant of $\phi^{\AA}(x)$ depends on the density $\nB$. In this way the ratio $\nA/\nB$ appears in the eigenvalue rescaling $\NA/2$ to larger or smaller values, according to the considered densities.

\begin{figure}
\begin{center}
\includegraphics[angle=-90,width=0.9\columnwidth]{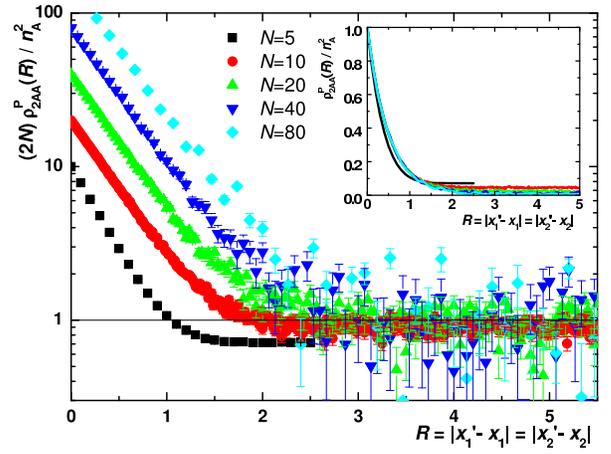}
\caption{Inset: projected two-body density matrix $\rho^P_{2\AA}(R)$, Eq.~(\ref{eq:PTBDM}), in a mixture of two ideal Bose gases for $\NA = \NB = 100; 80; 60; 40; 20; 10; 5$ (from upper to lower curves). Main figure: the same function multiplied by $2N$. }
\label{Fig:rho2l0(N)}
\end{center}
\end{figure}

In terms of the projected TBDM $\rho_2^P(R)$ (see Eq.~(\ref{eq:PTBDM})), the presence of ODLRO means a finite asymptotic $|R|\to\infty$ value. Contrary to dimensionless OBDM, where ODLRO is manifested as a constant independent of the number of particles, in dimensionless TBDM the finite value decreases with the number of particles as $1/N$. One way to understand this is that in $\rho_1(x,x')$ a particle is annihilated at $x$ and created at $x'$. The value of the wave function remains the same if this particle belongs to the condensate. In $\rho_2(x_1,x_2;x_1',x_2')$ a pair is displaced. In order to find a constant for a given particle $x_1$ from the molecular condensate the particle $x_2$ should belong to the same molecule. The probability of that is $1/(N-1)$. We account for this by multiplying $\rho^P_2(R)$ by $2N$ in the main figure~\ref{Fig:rho2l0(N)}, and a constant value is observed. The numerical error increases with the number of particles because the error is multiplied by $2N$, and in addition the phase space becomes larger and Monte Carlo simulation becomes more time consuming.
\begin{figure}
\begin{center}
\includegraphics[angle=-90,width=0.9\columnwidth]{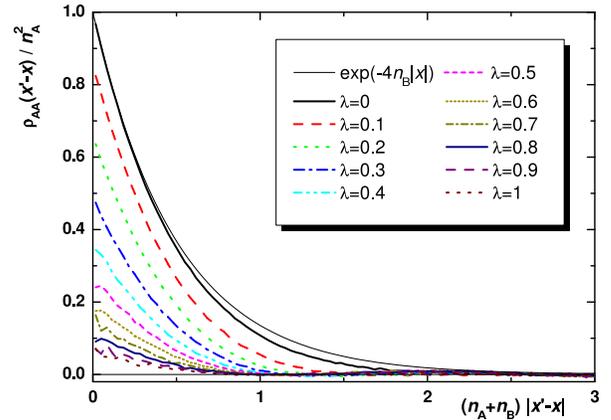}
\caption{Two-body density matrix $\rho_{2\AA}(r,L/4)$ in a balanced system of $\NA = \NB = 10$ particles for different interaction strengths $\lambda_\A = \lambda_\B = 0; 0.1; 0.2; 0.3; 0.4; 0.5; 0.6; 0.7; 0.8; 0.9; 1$ (decreasing from the value at zero). In the presence of BEC of molecules $\rho_{2\AA}(r,L/4)$ is proportional to the square of the molecular orbital. For $\lambda = 0$ the thermodynamic result $[\phi_{mol}^{\AA}(r)]^2$ (\ref{eq:phi_mol}) is shown for comparison.
}
\label{Fig:orbital}
\end{center}
\end{figure}

We have checked the dependence of the TBDM $\rho_2(R,r)$ on $R$, i.e. on the distance to which a pair is displaced. Its finite asymptotic value for $|R|\to\infty$ manifests the presence of ODLRO of molecules. For $|R|$ large enough, so that such an asymptotic value is reached, the shape of $\rho_2(R,r)$ in the $r$ direction corresponds to the square of the wave function of the bound state (molecular orbital) $[\phi_{mol}^{\AA}(r)]^2$. Fig.~\ref{Fig:orbital} shows the dependence on $r$ in a system of $N_\A = N_\B = 10$ particles for a value $R=L/4$ within the asymptotic regime. (For the dependence on $R$ refer to Fig.~\ref{Fig:rho2l0(N)}.)
Importantly, the range of the molecular orbital of an AA pair is governed by the density $\nB$ of the other component.

\begin{figure}
\begin{center}
\includegraphics[angle=-90,width=0.9\columnwidth]{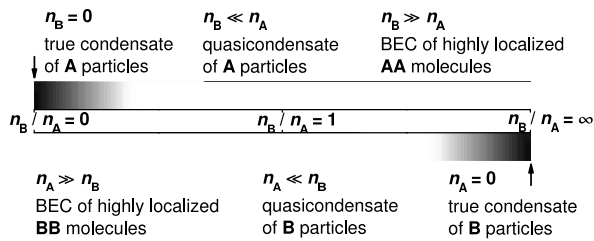}
\caption{Schematic phase diagram for $\lambda_\A = \lambda_\B = 0$ showing a crossover as the density imbalance $\nB/\nA$ is changed from zero to infinity. Component A passes through three phases 1: true BEC of A, 2: quasi-condensate of A, BEC of large AA dimers, 3: BEC of highly localized dimers (molecules). Component B passes through the same phases but in an inverted order. The system is normal in all regimes as no sound exits.
}
\label{Fig:diagram}
\end{center}
\end{figure}

The schematic phase diagram is drawn in Fig.~\ref{Fig:diagram} in terms of the density imbalance $\nB/\nA$. One way to follow it is to fix the number of A particles and change the number of particles of the other component. When B particles are absent the model reduces to an ideal Bose gas of A particles. Here a true Bose condensation of A particles is predicted. The transition to the phase with finite $\nB$ is not continuous which makes the case $\nB = 0$ an excluded point. Indeed, for any finite concentration of the other component, the asymptotic value of the OBDM~(\ref{eq:OBDM,l=0}) is strictly zero. Still, the decay could be extremely slow and the phase coherence might be preserved at distances large compared to the AA interparticle distance. We call this regime a ``quasi-condensate'' of A particles. Note that this notation differs from a quasicondensation in one-component systems (for example, Luttinger liquids), which in such systems is commonly understood as a slow power-law decay of OBDM. A finite density of the other component $\nB>0$ induces condensation of AA dimers. The dimers are shallow in the regime of quasicondensation of A. Instead for $\nB\gg \nA$ such dimers have extremely localized wave function (orbital) and we interpret this regime as a Bose-Einstein condensation of AA molecules.

The mapping preserves the excitation spectrum. From this we infer that the gas with $\lambda_\A=\lambda_B=0$ is not superfluid, as no sound exists and the system is not compressible.

\subsection{A mixture of Calogero-Sutherland gases with interspecies interaction: Fermi-momentum Bose condensate}

In this Section we discuss properties of a two-component gas with Calogero-Sutherland intraspecies potential and FTG interspecies interaction. The ground state wave function is given explicitly in Eq.~(\ref{eq:psi_0}). Although the Ideal Bose gas case is recovered in the limit $\lambda_\A = \lambda_\B = 0$, properties of the system at a small but finite value of $\lambda$ are qualitatively different. Indeed, the system becomes compressible due to interactions in each component. Appearance of the sound mode leads to superfluid behavior in a weakly interacting regime which has to be contrasted to a normal behavior of an ideal Bose gas.

The $p$-wave Feshbach resonance which leads to the FTG interaction causes exotic properties which are quite uncommon for a bosonic system. Indeed, it seems natural to expect that the ground-state wave function of a conventional bosonic system is positive-definite and it has no nodes. But it is not the case for our system. Indeed, the ground state wave function (\ref{eq:wfCS}) is symmetric under exchange of two A-A bosons or B-B bosons, according to bosonic statistics. As particles A and B are distinguishable, no symmetry or antisymmetry is required with the respect to A-B exchange. This means that both $s$-wave and $p$-wave scattering channels are available. Commonly $s$-wave scattering is considered, although in a vicinity of a $p$-wave resonance the relevant scattering corresponds to a locally antisymmetric solution. As a result the many-body wave function vanishes when two particles meet and also it changes sign.

The ``Feynman's no-node theorem'' \cite{Feynman} applies to almost all physical realizations of bosonic systems. The standard reasoning is that if the ground state solution $\Psi_0(x_1,...,x_N)$ has some nodes, its absolute value $|\Psi_0(x_1,...,x_N)|$ is also a solution. Smoothing the kinks of the solution close to the node leads to lowering the kinetic energy, while the potential energy increases only slightly. From this one concludes that the ground state wave function should be positive-definite and non-generate. This reasoning is not applicable in our case as the nodes in the wave function are due to the AB interaction potential and can not be removed. As a result the zero-temperature wave function is not positive definite and there are other solutions for the Hamiltonian that give the same ground state energy.

The possibility of bypassing Feynman's argument is already present in the literature. The proposals are to use the long-lived metastable states of bosons in the high orbital bands of optical lattices as a result of ``orbital-Hund's rule'' \cite{metastable} and multi-component bosons with spin-orbit coupling linearly dependent on momentum \cite{Wu08}. It is noted that the emergent states might experience unconventional Bose-Einstein condensation at non-zero momentum \cite{Wu09}. It is proposed that a mixture of two-species bosonic atoms interacting through a $p$-wave Feshbach resonance exhibits a finite-momentum atomic-molecular superfluid\cite{Radzihovsky09}.

Negative regions of the wave function will not change the properties of the local quantities (for example, AA pair distribution is the same as in a single component Calogero-Sutherland model), but dramatically modify non-local properties (for example, the one-body density matrix and the momentum distribution). In a single component gas with interaction parameter $\lambda$ the one-body density matrix has a power-law decay at large distances $\rho_{1}(x)\propto 1/|n x|^{\lambda/2}$ (all terms of the series expansions are explicitly obtained in Ref.~\cite{AGLS06} using replica method). For any finite value of $\lambda$ the asymptotic value of $\rho_1(x)$ vanishes and the ODLRO in one-particle density matrix is absent. Still the concept of a quasi-condensate can be applied in the regime of small $\lambda$, where the characteristic spreading of the one-body density matrix is large compared to the interparticle distance.

In the one-dimensional system some analogies between fermions and bosons with $p$-wave interaction can be drawn. By fixing positions of all particles but one we expect to see oscillations in the sign of the wave function as the particle is moved. In a Fermi system the sign changes each time the reference particle crosses another particle. In our system the sign changes each time A particle crosses B particle. The number of sign alterations is proportional to the density of the particles that produce change of the sign. In a fermionic system the characteristic period of space oscillations is fixed by the Fermi momentum, which itself is proportional to the density $k_\F = \pi n$. In our system the period of oscillations in one-body density matrix of A particles is fixed by the density of B particles with the corresponding momentum $k^B_F = \pi\nB$. We see that this characteristic momentum coincides with the Fermi momentum that the system $B$ would have if it were fermionic.

\begin{figure}
\begin{center}
\includegraphics[angle=-90,width=0.9\columnwidth]{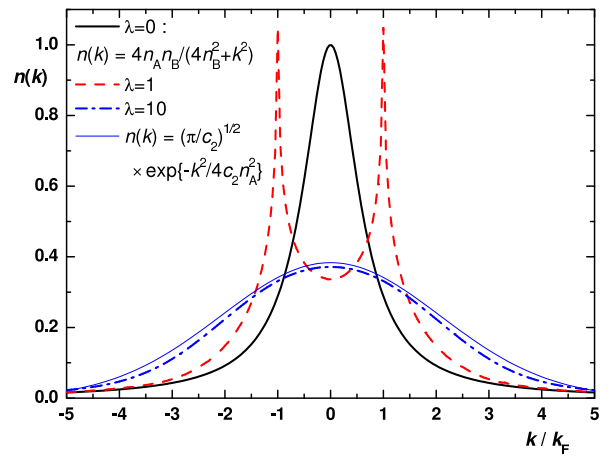}
\caption{Momentum distribution for a balanced system $\nA = \nB$ for characteristic values of the interaction parameter $\lambda = 0; 1; 10$. The solid line, $\lambda = 0$, shows the thermodynamic limit of a mixture of two ideal Bose gases (\ref{eq:nk,l=0}). Dashed line, $\lambda = 1$, and dash-dotted line, $\lambda = 10$ are results of Monte Carlo calculation with $N=40$ particles.
Thin solid line, Gaussian approximation to the momentum distribution in the ``classical" regime as given by Eq.~(\ref{nk:CLS}).
Note that the height of the peak for $\lambda = 1$ depends explicitly on the number of particles in the system.}
\label{Fig:nk}
\end{center}
\end{figure}

The momentum distribution $\nA(k)$ is shown for some characteristic values of $\lambda$ in Fig.~\ref{Fig:nk}. For a mixture of two ideal Bose gases the momentum distribution is a Lorentzian (\ref{eq:nk,l=0}) with the maximum at zero momentum. In this case the value at the peak is equal to the density of the corresponding component, i.e. $\nA(k) = \nA$. For $\lambda = 1$ the maximum is shifted to the Fermi momentum of the other component $k = k^B_F$. For very large values of $\lambda$ there is again only one peak situated at zero momentum.
In this regime the potential is much larger then the kinetic energy which leads to creation of a quasicrystal\cite{AGLS06}. The momentum distribution can be approximated by a Gaussian, similarly to classical systems. The Gaussian width can be extracted from the short-range series expansion of the one-body density matrix $g_{1\A}(x) = 1 - c_2 (\nA x)^2+...$ by approximating it with a Gaussian $g_{1\A}(x) = \exp\{-c_2 (\nA x)^2\}$ \cite{Mazzanti08} and calculating its Fourier transform
\begin{eqnarray}\label{nk:CLS}
n(k) = \sqrt{\frac{\pi}{c_2}}\exp\left\{-\frac{k^2}{4c_2\nA^2}\right\}\ .
\end{eqnarray}
The coefficient $c_2$ is related to the kinetic energy, which itself can be obtained by using the Hellman-Feynman theorem from the total energy.  The resulting expression is $c_2 = \lambda^2/(6(2\lambda-1))$ \cite{AGLS06}.

\begin{figure}
\begin{center}
\includegraphics[angle=0,width=0.9\columnwidth]{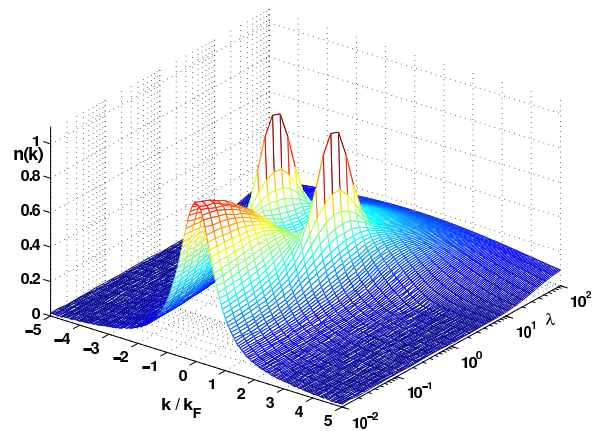}
\caption{3D plot of the momentum distribution for a balanced system $\nA = \nB$ as a function of the momentum and the interaction parameter $\lambda$.}
\label{Fig:nk3D}
\end{center}
\end{figure}

The large-$k$ asymptotic of the momentum distribution is governed by the short-range physics and it can be obtained from the short-range behavior of the two-body Bijl-Jastrow terms. The Calogero-Sutherland interaction introduces a cusp when two particles meet each other as described by $|\sin(\pi(x_i-x_j))/L|^\lambda$ term in the ground state wave function. The momentum distribution can be written in the following form:
\begin{equation}
n(k) = n \int |\Psi_0(k_1,x_2,...)|^2 dx_2...dx_N\ ,
\label{eq:n(k)=|psi|^2}
\end{equation}
where we use the momentum representation of the ground state wave function with respect to the first particle\cite{Olshanii03}
\begin{equation}
\Psi_0(k_1,...) = \int_{-L/2}^{L/2} e^{-i k_1 x_1}\Psi_0(x_1,...)\;dx_1\ .
\label{eq:Psi(k)}
\end{equation}
Here for simplicity we write down only the first argument, i.e. $\Psi_0(k_1,...) = \Psi_0(k_1,x_2,...,x_\NA;y_1,...,y_\NB)$. First we consider the case of the one-component Calogero-Sutherland gas, $\NB = 0$. For small distances between two particles the sine function in (\ref{eq:psi_0}) can be expanded to $\pi|(x_i-x_j)/L|^\lambda$ with the relevant part of the integral being proportional to
\begin{equation}
\Psi_0(k)\propto\int e^{-i k x}|x|^\lambda dx\ .
\label{eq:Psi(k)int}
\end{equation}
As the relevant part of the integral is coming from small $x$ we are free to change the limits of the integration and make them infinite. In order to make the resulting integral convergent we multiply the integrant by an exponent $\exp(-a|x|)$, evaluate the integral and then take the limit of $a\to 0$. This results in a power-law dependence $\Psi_0(k,...)\propto 1/|k|^{1+\lambda}$. The momentum distribution (\ref{eq:n(k)=|psi|^2}) then follows a power-law $n(k) \propto 1/|k|^{2(1+\lambda)}$ with the exponent depending on the value of $\lambda$. In a two-component system, $\NB>0$, the integrand in (\ref{eq:Psi(k)}) changes its sign each time $x$ passes through a particle of the other type $y_i$. This splits the integral (\ref{eq:Psi(k)int}) into different regions of integration with the leading contribution from the each point $y_i$ being of the order of $1/|k|$. This changes the asymptotic behavior of the momentum distribution behavior to $n(k) \propto 1/k^2$ for all physical values of $\lambda\ge 0$. As already noted above, the momentum distribution in the $\lambda=0$ case can be explicitly evaluated as given by  Eq.~(\ref{eq:nk,l=0}) and it has a $1/k^2$ tail. The case of $\lambda = 1$ corresponds to the TG interaction potential with the Bijl-Jastrow term proportional to the absolute value of the interparticle distance $|x_i-x_j|$. Similarly the Lieb-Liniger gas has a two-particle solution of $|x_i-x_j-a_{1\D}|$, where $a_{1\D}$ is the one-dimensional $s$-wave scattering length. From this we conclude that the asymptotic of the momentum distribution of TG and Lieb-Linger gases are $n(k)\propto 1/|k|^4$ for a single component system and $n(k)\propto 1/|k|^2$ for a two-component system with FTG interspecies potential.

\begin{figure}
\begin{center}
\includegraphics[angle=-90,width=\columnwidth]{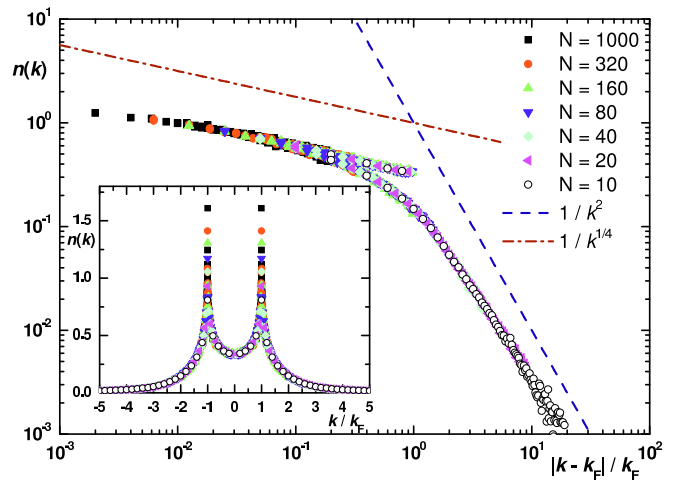}
\caption{Inset: Momentum distribution of a balanced system with $\lambda_\A=\lambda_\B=1$ for different system sizes $N_\A = N_\B = 10; 20; 40; 80; 160; 320; 1000$ as a function of momentum in units of Fermi momentum $k/k_\F$. Main figure: the same data points shown on a log-log scale as a function of the separation from the divergency point $|k-k_\F|/k_\F$. High-momentum $1/k^2$ and low-momentum $1/|k|^{1/4}$ asymptotics are shown for comparison.}
\label{Fig:nk(kF)}
\end{center}
\end{figure}
An example of finite-size dependence of the momentum distribution is shown in Fig.~\ref{Fig:nk(kF)}. The behavior of the tail is best seen on a logarithmic scale which we adopt in the main plot. The analytic prediction $1/k^2$ is shown with a dashed line which on the  logarithmic scale is parallel to the numerical curve, while the constant shift is related to the proportionality constant $C_{asympt}$ in $n(k) = C_{asympt}/k^2$.

The Fermi-momentum condensation peak is the most pronounced for values of $\lambda=1$ as can be seen from Fig.~\ref{Fig:nk3D}. We report the finite-size dependence of momentum distribution exactly for this value of $\lambda$ in Fig.~\ref{Fig:nk(kF)}. The one-dimensional analogue of Bose-Einstein condensation at some value of momentum $k$ is understood as divergency of the momentum distribution at this momentum $n(k)\to\infty$ in the thermodynamic limit $N\to\infty$. The momentum distribution for different number of particles is shown on a linear scale in the instet of Fig.~\ref{Fig:nk3D}. One sees that there are high peaks at values of momenta $k=\pm k_\F$ and the height of the peaks increases with the number of particles. In order to find the functional form of the divergency it is convenient to plot the momentum distribution as a function of $|k-k_\F|$ on a log-log scale, as shown in the main plot of Fig.~\ref{Fig:nk3D}. We see that the divergency is well described by a power-law $n(k)\propto 1/|k|^{1/4}$ and the height of the peak increases as $N^{1/4}$. This should be contrasted with the infrared divergency of a one-component Tonks-Girardeau gas of $N^{1/2}$.

\begin{figure}
\begin{center}
\includegraphics[angle=0,width=\columnwidth]{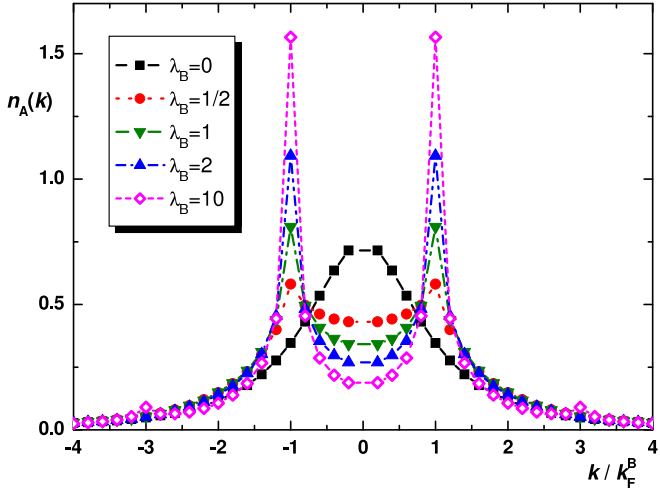}
\caption{Momentum distribution of a system with $N_\A=N_\B=10$ particles with the AA interaction strength fixed to $\lambda_\A=1$ for different interactions strengths of the BB component $\lambda_\B=0; 1/2; 1; 2; 10$ as a function of momentum in units of Fermi momentum $k/k_\F$. Compare the data to the inset of Fig.~\ref{Fig:nk(kF)}.
}
\label{Fig:nklambdaB}
\end{center}
\end{figure}

For the $\lambda_\A=1$ case which leads to the highest peak in $\nA(k)$ at the Fermi momentum, we study the dependence on the BB interaction strength $\lambda_\B$. The results are shown in Fig.~\ref{Fig:nklambdaB}. When the other component is an ideal Bose gas, $\lambda_\B=0$, the positions of $B$ particles are uncorrelated and the one-body density matrix $g_{1\A}(x)$ decays in a monotonic way to zero. In this case the peak in the Fermi momentum is absent. The oscillations in $g_{1\A}(x)$ are enhanced by ordering in the B component, which appears in the system as $\lambda_\B$ is increased. In particular, the height of the peak in $\nA(k)$ at the Fermi momentum is  greatly increased when the other component forms a quasicrystal (see $\lambda_\B=10$ case in Fig.~\ref{Fig:nklambdaB}).

\subsection{A mixture of Calogero-Sutherland gases with interspecies interaction: off-diagonal properties}

Examples of the typical behavior of the one-body density matrix in a balanced system $\nA = \nB$ are shown in Fig.~\ref{Fig:rho1(l)} for different values of interaction parameter $\lambda$. One sees how the fermionic nature of $\A \B$ interactions manifests in oscillating behavior of $\rho_1(x-x') = \langle\Psi^\dagger(x)\Psi(x')\rangle$. While removing a particle from $x$ and moving it to $x'$, the ground state wave function changes its sign each time $x'$ crosses a particle of the other component. In the non-interacting case, $\lambda=0$, there are no correlations between particle positions and for two particles there is a linear decay $\rho_{1\A}(x) = \nA(1-2\nB|x|/\NB)$. In the thermodynamic limit the decay in $\rho_{1\A}(x)$ is exponentially fast according to Eq.~(\ref{eq:OBDM,l=0}), but still the one-particle density matrix remains positive. Instead, for a finite $\lambda$ the interparticle correlations are nontrivial and the sign alternations due to FTG interactions lead to regions of negative sign.

\begin{figure}
\begin{center}
\includegraphics[angle=-90,width=0.9\columnwidth]{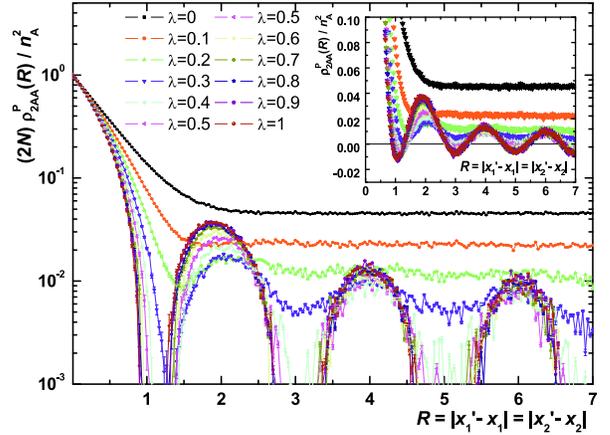}
\caption{Inset: projected two-body density matrix $\rho^P_{2\AA}(R)$, Eq.~(\ref{eq:PTBDM}) for $\NA = \NB = 10$.  $\lambda_\A = \lambda_\B = 0; 0.1; 0.2; 0.3; 0.4; 0.5; 0.6; 0.7; 0.8; 0.9; 1$ (from upper to lower curves). Main figure: logarithmic scale. Inset: close up of the tail.}
\label{Fig:rho2(l)}
\end{center}
\end{figure}

We study the ODLRO in the two-body density matrix. Fig.~\ref{Fig:rho2(l)} shows its behavior on the displacement distance $R$ in a system with $N_\A = N_\B = 10$ particles and different values of interaction parameter. For $\lambda=0$ the function goes to a constant value as $|R|\to\infty$ ($\rho_2(R)\to 1/20$ in the present case). The ODLRO gradually vanishes as the interaction parameter is increased and at a certain point the oscillatory behavior dominates (see inset of Fig.~\ref{Fig:rho2(l)}). In the regime where the ODLRO is still present we study the shape of the dimer wave function (orbital). We show the cut of the two-body density matrix for the displacement $R = |x_1-x_1'|=|x_2-x_2'|=L/4$ as a function of the dimer size $|x_2-x_1|$ in Fig.~\ref{Fig:orbital}. For small values of $\lambda$ the function is significantly different from zero and the decay is exponentially fast.
According to the spectral decomposition~(\ref{eq:spectral_decomposition}) the function shown in Fig.~\ref{Fig:orbital} is proportional to the the square of the molecular orbital. Of course, this relation has no meaning when ODLRO is absent and $\rho_2(r,L/2)$ can become negative for large $\lambda$.

\subsection{A mixture of Calogero-Sutherland gases with interspecies interaction: phase diagram}

In order to understand if the system is superfluid we note that the mapping applies also to the excited states. The excitation spectrum in a single component Calogero-Sutherland model for small frequencies $\omega$ can be analyzed from the Luttinger liquid theory. The dynamic form factor $\sigma(\omega,k)$ in the point where the excitation spectrum touches zero, $k=2\pi n$, has a power-law dependence on the frequency $\sigma(\omega,2\pi n)\propto \omega^{2/\lambda-2}$ for small $\omega$ \cite{Castro,AP04}. In the weakly interacting regime (i.e. small $\lambda$) the weight is vanishingly small and the behavior of the system is analogous a three-dimensional superfluid. Instead, as $\lambda$ is increased the system starts behaving as a normal system and an impurity dragged through the system will cause an energy dissipation \cite{AP04}.

\begin{figure}
\begin{center}
\includegraphics[angle=0,width=\columnwidth]{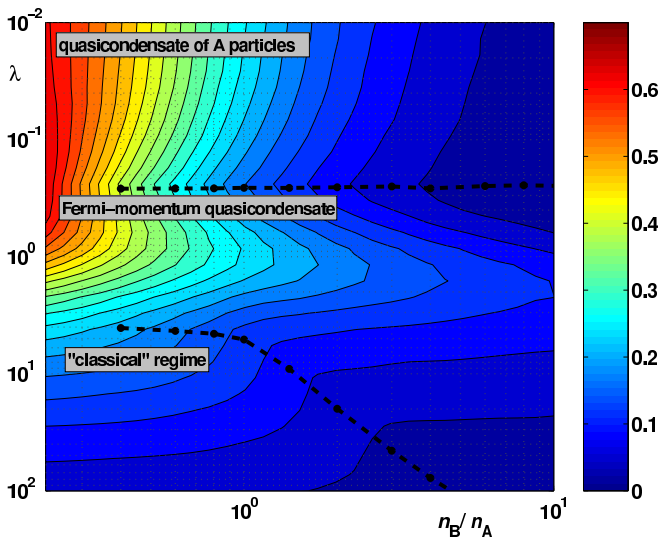}
\caption{Altitude of the highest peak in the momentum distribution $\max\limits_{k}n(k)$ for $N_\A=10$ and $N_\B=2$---$100$ particles shown as a contour plot in variables of the density imbalance $\nB/\nA$ and interaction parameter $\lambda$. The dashed lines separate different phases and correspond to parameters for which the height of the peaks at $k=0$ and $k=k_\F$ is equal. Some important features of the phase diagram are observed in this plot. True Bose-Einstein condensation (constant ODLRO in $\rho_1(x)$) of A (B) particles happens for $\lambda=0$ and $\nB=0$ ($\nA=0$). Small, but finite values of $\lambda$ and $\nB$ ($\nA$) correspond to quasicondensate of A (B) particles with the highest peak positioned at $k\approx 0$. By increasing $\lambda$ the position of the highest peak shifts to $k = k_\F$ which corresponds to the phase of Fermi-momentum condensation. The strongest divergence at the Fermi momentum is observed for
$\lambda = 1$. By increasing the value of $\lambda$ further the position of the highest peak shifts once more to $k\approx 0$ which marks the ``classical'' regime.
We stop the dashed line for small $\nB$ when $k_\F^\B$ is comparable to the momentum quantization $2\pi/L$.
In the ideal gas limit $\nB\to 0$ the Fermi momentum vanishes as $k_\F^\B\to 0$.
}
\label{Fig:diagram2}
\end{center}
\end{figure}

We summarize the phase diagram of a system with $\lambda_\A = \lambda_\B$ in Fig.~\ref{Fig:diagram2}. The contour plot shows the maximal value of the momentum distribution $n_{\A}(k)$ as a function of $\lambda$ and density imbalance $\nB/\nA$. In the noninteracting limit $\lambda\to 0$ we recover the phase diagram shown in Fig.~\ref{Fig:diagram}. The maximum observed for small density of the other component $\nB/\nA\ll 1$ corresponds to a very high peak of the momentum distribution at $k=0$, which we associate with a quasicondensate of A particles.
At the same time in this regime there is a true condensation of AA pairs (molecules).
The quasicondensate regime extends as well to finite values of $\lambda$ where it is limited by a power-law decay in terms of the $\lambda$ variable and by an exponential decay in terms of $\nB/\nA$. A further increase in $\lambda$ leads to stronger interparticle correlations and around $\lambda=1$ we find an appearance of another maximum, which this time is situated at the Fermi momentum $k=k_\F$. We associate this regime with the Fermi-momentum quasicondensation. In order to estimate a border line between the different phases we plot with black circles the points on the phase diagram where the height of the peak at $k=0$ is equal to the height of the peak at $k=k_\F$. For even stronger correlations, $\lambda\gg 1$, the position of the peak shifts again to $k=0$ (as an example momentum distribution for $\nA=\nB$ in Figs. \ref{Fig:nk} and \ref{Fig:nk3D}). In this regime the interactions quickly destroy the coherence (see $\lambda=10$ case in Fig.~\ref{Fig:rho1(l)}) and from this point of view the systems behaves quite similarly to a classical one. The kinetic energy is much smaller than the repulsion between the particles and this leads to creation of a crystal-like structure. The momentum distribution can be closely described by a Gaussian with the peak at $k=0$, Eq.~(\ref{nk:CLS}). We refer to this regime as a ``classical'' phase.

It is important to note that there are no phase transitions in our system. Instead the dashed lines in Fig.~\ref{Fig:diagram2} separate different physical regimes and the transition between them is continuous (crossover). Furthermore, the position of the lines depends explicitly on the number of particles. This can be understood by noticing that divergence at $k=k_\F$ in a Fermi quasicondensate is weaker compared to the divergence at $k=0$ in a quasicondensate; see Fig.~\ref{Fig:nk(kF)}.

As discussed above, the system is normal for the value $\lambda=0$ and arbitrary densities $\nA/\nB$, which is an excluded line. Instead, for finite but small values of $\lambda$ the system behaves practically as a superfluid, while at larger values of $\lambda$ there is a strong friction in the system and its behavior is normal. In the limit of large $\lambda$ the interparticle correlations are extremely strong, which is a one-dimensional analogue of a crystal. In this regime the system is normal and no Bose condensation of any kind is present.

\section{Conclusions}

To summarize, we have studied properties of a bosonic two-component one-dimensional system at zero temperature. The interactions between particles of the same component (AA or BB) are taken to be of a Calogero-Sutherland type $\lambda(\lambda-1)/(x_i-x_j)^2$. The interaction parameters $\lambda_\A$ and $\lambda_\B$ can be changed independently and are related in a simple way to the Luttinger parameters $K_\A = 1/\lambda_\A$, $K_\B = 1/\lambda_\B$. In this way the long-range properties of the considered model are universal for Luttinger liquids. For the AB interaction we consider $p$-wave scattering channel with attractive interaction such that the wave function of a two body zero-energy scattering solution is flat outside of some short range, i.e., the usual fermionic Tonks-Girardeau (FTG) potential. We write explicitly the many-body wave function which we obtain by a mapping procedure. The same mapping applies to the excited states of the system. The phase diagram of the system is very rich and is governed by the interaction strengths $\lambda_\A, \lambda_\B$ and the density ratio $N_\A/N_\B$. The limiting case of $\lambda_\A=\lambda_\B=0$ corresponds to a mixture of two ideal Bose gases with FTG interaction between them and was recently studied in Ref.~\cite{Gir09_3}. In this case a single component $N_\B=0$ ideal gas creates a completely ``true'' Bose-Einstein condensate with all particles being in the same state. Presence of the other component destroys the off-diagonal long-range order in the one-body density matrix. At the same time there is a divergent peak in the momentum distribution at $k=0$, which we interpret as a quasi-condensation. The one-body density matrix, two-body density matrix and the momentum distribution of such a system can be obtained analytically. There is a full condensation of AA pairs (molecules) as manifested by the presence of the off-diagonal long range order in the two-body density matrix, which for large displacement of a pair factorizes to the product of molecular orbitals.
There is a BEC-BCS-like crossover from AA-pair BEC when $\nB\gg\nA$, implying tightly bound AA pairs, to extended and strongly overlapping AA Cooper pairs when $\nB\ll\nA$.

The properties of the system with finite values of $\lambda$ are obtained numerically by sampling the known ground state wave function with the Monte Carlo technique.
The interactions between the particles of the same species destroy the ODLRO both in one- and two- body density matrices. Still, the regime of quasicondensation of A particles persists for small interaction strength $\lambda$ and small concentration of the other component $\nB/\nA$. The $p$-wave interactions leads to appearance of fermionic-like properties, such as oscillations around zero in the one-body density matrix, which leads to a divergence in the momentum distribution $\nA(k)$ at the Fermi momentum of the other component $k=\pm k_\F^\B=\pm\pi\nB$. We refer to this regime as a phase of Fermi-momentum quasicondensation.

By making the interactions with the species stronger we reach the regime where the potential energy of repulsive interaction is much larger than the kinetic energy which leads to appearance of a crystal-like order. We refer to this regime as  ``classical''. There is a very rapid decay of coherence in the one-body density matrix. The momentum distribution is well approximated by a Gaussian, for which we provide an explicit expression.

Finally, superfluid properties of the system are discussed on the basis of the Luttinger liquid approach.

\begin{acknowledgments}
This work was initiated at the Aspen Center for Physics during the summer 2009 workshop ``Quantum Simulation/Computation with Cold Atoms and Molecules''. We are grateful to the organizers Lincoln Carr, Erich Mueller, Ignacio Cirac, and David Weiss for the opportunity to participate, and to Peter Reynolds for helpful comments. The Aspen Center for Physics is supported by the U.S. National Science Foundation, research of M.D.G. by the U.S. Army Research Laboratory and the U.S. Army Research Office under grant number W911NF-09-1-0228, and that of G.E.A. is supported by post doctoral fellowship by MEC (Spain), (Spain) Grant No. FIS2008-04403.

\end{acknowledgments}

\begin{thebibliography}{3}
%
\bibitem{Ols98} M. Olshanii, \Journal{\PRL}{81}{938}{1998}.
%
\bibitem{GraBlu04} B.E. Granger and D. Blume, \Journal{\PRL}{92}{133202}{2004}.
%
\bibitem{Rob01} J.L. Roberts {\it et al.}, \Journal{\PRL}{86}{4211}{2001}.
%
\bibitem{Par04Kin04} B. Paredes, {\it et al.}, Nature {\bf 429}, 277 (2004); T. Kinoshita, T.R. Wenger, and D.S. Weiss, \Journal{\Science}{305}{1125}{2004}.
%
\bibitem{Kin05} T. Kinoshita, T.R. Wenger, and D.S. Weiss, \Journal{\PRL}{95}{190406}{2005}.
%
\bibitem{Kin06} T. Kinoshita, T.R. Wenger, and D.S. Weiss, Nature {\bf 440}, 900 (2006).
%
\bibitem{Gir60Gir65} M. Girardeau, \Journal{\JMP}{1}{516}{1960};
M.D. Girardeau, \Journal{\PR}{139}{B500}{1965}, Secs. 2, 3, and 6.
%
\bibitem{GirOls04} M.D. Girardeau and M. Olshanii, \Journal{\PRA}{70}{023608}{2004}.
%
\bibitem{GirNguOls04} M.D. Girardeau, Hieu Nguyen, and M. Olshanii, Optics Communications {\bf 243}, 3 (2004).
%
\bibitem{Tik04} C. Ticknor, C.A. Regal, D.S. Jin, and J.L. Bohn, \Journal{\PRA}{69}{042712}{2004}.
%
\bibitem{Gir09_3} M.D. Girardeau, \Journal{\PRL}{102}{245303}{2009}.
%
\bibitem{LieLin63} E.H. Lieb and W. Liniger, Phys. Rev. {\bf 130}, 1605 (1963).
%
\bibitem{C69S71} F. Calogero, \Journal{\JMP}{10}{2191, 2197}{1969};
B. Sutherland, \Journal{\JMP}{12}{245}{1971}.
%
\bibitem{Note1} The TG gas must be understood not as the result of simply substituting $\lambda=1$ in (\ref{eq:wfCS}), but rather as
the consequence of two noncommuting limits $\lambda\to 1+$ and $x\to 0$. Substituting $\lambda=1+\epsilon$ with $\epsilon$ a positive
infinitesimal gives $V^{\CS}(x)=\epsilon/x^2+\mathcal{O}(\epsilon^2)$. Hence $V^{\CS}(x)\to 0$ as $\epsilon\to 0$ for $x\ne 0$, but $V^{\CS}(x)\to +\infty$ as $x\to 0$ for $\epsilon>0$. In this way $V^{\CS}(x)$ with $\lambda=1+$ simulates the point hard core interaction of the TG gas. For this and other subtleties of the CS model see, e.g., E. Langmann,
http://www.emis.de/journals/SIGMA/2007/031/sigma07-031.pdf.
%
\bibitem{Reatto} L. Reatto and G.V. Chester, Phys. Rev. {\bf 155}, 88 (1967), particularly Sec. VI, Eq. (6.2).
%
\bibitem{AGLS06} G.E. Astrakharchik, D.M. Gangardt, Yu. E. Lozovik, and I.A. Sorokin, \Journal{\PRE}{74}{021105}{2006}.
%
\bibitem{Landau} L.D. Landau and E.M. Lifshitz, \textit{Quantum Mechanics, Nonrelativistic Theory} (Pergamon Press, London, 1958)
%
\bibitem{Note1.5} See the paragraph following Eqs. (44-46) of \cite{AGLS06}.
%
\bibitem{Note2} By the same argument as in \cite{Gir09_3}, both $\NA$ and $\NB$ must be even to ensure periodicity
in all A-particle and B-particle coordinates.
%
\bibitem{GirMin06} M.D. Girardeau and A. Minguzzi, \Journal{\PRL}{96}{080404}{2006}.
%
\bibitem{Metropolis53} N. Metropolis, A. W. Rosenbluth, M. N. Rosenbluth, A. H. Teller, and E. Teller, J. Chem. Phys., {\bf 21}, 1087 (1953)
%
\bibitem{Senatore} S. De Palo, F. Rapisarda and G. Senatore, Phys. Rev. Lett. {\bf 88}, 206401 (2002).
%
\bibitem{Ortiz} G. Ortiz and J. Dukelsky, cond-mat/0503664; L. Salasnich, N. Manini and A. Parola, cond-mat/0506074.
%
\bibitem{ABCG05} G. E. Astrakharchik, J. Boronat, J. Casulleras, and S. Giorgini, Phys. Rev. Lett. {\bf 95}, 230405 (2005)
%
\bibitem{Mazzanti08} F. Mazzanti, G. E. Astrakharchik, J. Boronat, and J. Casulleras,
Phys. Rev. A {\bf 77}, 043632 (2008)
%
\bibitem{Castro}  A. N.~Castro Neto et al., Phys. Rev. B {\bf 50}, 14032 (1994).
%
\bibitem{AP04} G.E. Astracharchik and L.P. Pitaevskii, Phys. Rev. A {\bf 70}, 013608 (2004).

\bibitem{Feynman} R. P. Feynman, Statistical Mechanics, A Set of Lectures (Addison-Wesley Publishing Company, 1972).

\bibitem{metastable} W. V. Liu and C. Wu, Phys. Rev. A {\bf 74}, 13607 (2006), C. Wu, W. V. Liu, J. E. Moore, and S. Das Sarma, Phys. Rev. Lett. {\bf 97}, 190406 (2006), V. M. Stojanovic, C. Wu, W. V. Liu, and S. D. Sarma, Phys. Rev. Lett. {\bf 101}, 125301 (2008).

\bibitem{Wu08} C. Wu and I. M. Shem, 
arXiv:0809.3532 (2008).

\bibitem{Wu09}  C. Wu, 
Mod. Phys. Lett. B {\bf 23}, 1 (2009)

\bibitem{Radzihovsky09} L. Radzihovsky and S. Choi, 
Phys. Rev. Lett. {\bf 103}, 095302 (2009)

\bibitem{Olshanii03} M. Olshanii and V. Dunjko 
Phys. Rev. Lett. {\bf 91}, 090401 (2003)


\end{thebibliography}
\end{document}